\documentclass[%
 aip,
 amsmath,amssymb,
reprint, onecolumn 
]{revtex4-1}

\usepackage{graphicx}
\usepackage{dcolumn}
\usepackage{bm}

\usepackage[utf8]{inputenc}
\usepackage[T1]{fontenc}
\usepackage{mathptmx}
\usepackage{etoolbox}
\usepackage{graphicx}
\usepackage{float} 
\makeatletter
\def\@email#1#2{%
 \endgroup
 \patchcmd{\titleblock@produce}
  {\frontmatter@RRAPformat}
  {\frontmatter@RRAPformat{\produce@RRAP{*#1\href{mailto:#2}{#2}}}\frontmatter@RRAPformat}
  {}{}
}%
\makeatother

\usepackage{amsmath,amssymb,graphicx}
\usepackage{amsthm}
\usepackage{derivative}

\theoremstyle{definition}

\theoremstyle{remark}

\renewcommand{\thesection}{\arabic{section}}
\renewcommand{\thesubsection}{\arabic{section}.\arabic{subsection}}
\renewcommand{\thesubsubsection}{\arabic{section}.\arabic{subsection}.\arabic{subsubsection}}

\numberwithin{equation}{section}

\usepackage{hyperref}

\begin{document}


\title[]{ PT-Symmetry in
$2\times 2$ Matrix Polynomials Formed by Pauli Matrices}
\author{Stalin Abraham}%
\email{stalinabraham.v@gmail.com}%
\affiliation{%
    \textsuperscript{1}School of Physical Sciences, UM-DAE Centre for Excellence in Basic Sciences, University of Mumbai, Kalina Campus, Mumbai 400 098, India
}

\author{Ameeya A. Bhagwat}%
\affiliation{\textsuperscript{1}School of Physical Sciences, UM-DAE Centre for Excellence in Basic Sciences, University of Mumbai, Kalina Campus, Mumbai 400 098, India
}

\date{\today}

\begin{abstract}
\textbf{Abstract:}
$2\times2$ matrix polynomials of the form $P_{n}(z)= \sum^{n}_{j=0}\,\sigma_{j}\,z^{j}$, for the cases $n=1,2,3$ are constructed, and the nature of PT-symmetry is examined across different points $z=(x,y)$ in the complex plane. The PT-symmetric properties of $P_{n}(z)$ can be characterized by two functions, denoted by $s(x,y)$ and $h(x,y)$. If the trace of the matrix polynomial is real, then the points at which it can exhibit PT-symmetry are defined by the family of curves $s(x,y)=0$. Additionally, at points where the function $h(x,y)\geq 0$, the matrix polynomial exhibits unbroken PT-symmetry; otherwise, it exhibits broken PT-symmetry. The intersection points of the curves $s(x,y)=0$ and $h(x,y)=k$, for a given $k\in \mathbb{R}$, are shown to lie on an ellipse, hyperbola, two lines passing through the origin, or a straight line, depending on the nature of PT-symmetry of the matrix polynomial. The PT-symmetric behaviour of $P_{n}(z)$ at the zeros of the matrix polynomial is also studied.

\end{abstract}

\maketitle

\section{Introduction}
PT-symmetric and Pseudo Hermitian operators represent a class of operators that, while generally non-Hermitian, can still yield a real spectrum. Such operators are important in the field of non-Hermitian quantum mechanics and may serve as the Hamiltonians for various quantum mechanical systems. This perspective allows us to look beyond the traditional Hermitian framework of quantum mechanics, which has led to the development of non-hermitian quantum mechanics and its applications in several areas of physics.

Towards the end of the 1990s, Bender and collaborators demonstrated that non-Hermitian operators could have a real spectrum by relaxing the requirement for observables to be Hermitian and instead emphasizing parity-time reversal (PT) symmetry\cite{Bender:1998ke, Bender:1998gh, Bender:2002vv}. 
If all the eigenstates of a PT-symmetric operator are also eigenstates of the PT operator, then all of its eigenvalues are real. 
Otherwise,  the PT-symmetric operator has at least one complex conjugate pair of eigenvalues \cite{Bender:2003gu}, thereby allowing one to describe systems with dissipation\cite{Wang201322}. The former scenario is referred to as unbroken PT-symmetry, while the latter is referred to as broken PT-symmetry \cite{Bender:2003gu}.

Ali Mostafazadeh has studied a specific class of non-Hermitian operators known as pseudo-Hermitian operators\cite{Mostafazadeh:2002id,Mostafazadeh:2001nr,Mostafazadeh:2001jk}, and he showed that all diagonalizable PT-symmetric Hamiltonians are pseudo-Hermitian \cite{Mostafazadeh:2002id}.
The necessary and sufficient conditions for PT-symmetric and non-Hermitian Hamiltonians to exhibit real spectra have been investigated by Bender and Mostafazadeh~\cite{Mostafazadeh:2001jk, Bender:2009mq}. 
Ruili Zhang and colleagues demonstrated that, in finite-dimensional cases, a PT-symmetric Hamiltonian is always pseudo-Hermitian, irrespective of its diagonalizability \cite{Zhang:2019gyc}. 
Additionally, Qing-hai Wang established the equivalence between PT-symmetric and pseudo-Hermitian matrices in the context of two-level systems \cite{Wang201322}.
The research on non-Hermitian Hamiltonians continues to be a significant and active area of study~\cite{ruter2010observation, Schindler:2011mnw, Bender:2012je, bittner2012pt, zhang2016observation, tommy2020,abraham2024classificationgeometriccharacterizationsensembled}. 
 
 Building on the concept of PT-symmetry, we explore matrix polynomials that exhibit this symmetry. Such an investigation may demonstrate how the matrix polynomials' properties are related to the system's behaviour with PT-symmetry. 
 A matrix polynomial $P$ is a polynomial in a complex variable with matrix coefficients~\cite{SINAP199627}.
 In this article, we constructed $2\times2$ matrix polynomials of the form $P_{n}(z)= \Sigma^{n}_{j=0}\,\sigma_{j}\,z^{j}$, for the cases $n=1,2,3$, where $n$ is the degree of the matrix polynomial, $\sigma_0$ is the identity matrix and $\sigma_{j}$,($j=1,2,3)$ are the Pauli matrices.
 The matrix polynomials constructed in this way are not PT-symmetric in general.
 Therefore, we investigated the question of which values of $z \in\mathbb{C}$, the matrix polynomials, could exhibit PT-symmetry.
 Since the trace of $P_{n}(z)$ is real, the PT-symmetric nature of the matrix polynomial can be analyzed by two functions associated with the determinant of $P_{n}(z)$, given by $s(x,y)$ and $h(x,y)$.
 A $2\times2$ matrix polynomial with real trace exhibits PT-symmetry at all points $z \in\mathbb{C}$, defined by the family of curves $s(x,y)=0$. Furthermore, at all points in the complex plane where the family of curves defined by $s(x,y)=0$ intersects with the family of curves $h(x,y)=k$, such that $k\geq0$, the matrix polynomial has unbroken PT-symmetry. Conversely, if $k<0$, it exhibits broken PT-symmetry.

 The matrix polynomials $P_{n}(z)$ (for $n=1,2,3$) become Hermitian for all $z$ lying on the x-axis and for all other $z$ where $P_{n}(z)$ (for $n=1,2,3$) possesses PT-symmetry, it is non-Hermitian.
 The families of curves associated with the matrix polynomial $P_{n}(z)$ (for n=1,2,3), defined by the equations $s(x,y)=0$ and $h(x,y)=k$ for a given $k \in\mathbb{R}$, intersect at most $2n$ points in the complex plane(note that this is not true if the functions  $s(x,y)$ and $h(x,y)$ are constant functions). In section 2, We have shown that for some specific values of $k$, such intersection points are minimal in number, resulting in the fewest points in the complex plane where $P_{n}(z)$ exhibits PT-symmetry. Further, for the case of $P_{1}(z)$ ($n=1$), the intersection points of the families of curves $s(x,y)=0$ and $h(x,y)=k$ for a given $k\neq0$ are observed to lie on a hyperbola. When $k=0$, the intersection point lies on two straight lines passing through the origin. 
 In the case of $P_{2}(z)$ ($n=2$), the intersection points of the families of curves $s(x,y)=0$ and $h(x,y)=k$  lie on a circle for a given $k<0$, and the intersections points lie on an ellipse for a given $k>0$. When $k=0$, the intersection points of the families of curves $s(x,y)=0$ and $h(x,y)=k$ are shown to lie on a straight line (y-axis).
 In the case of $P_{3}(z)$ ($n=3$), the intersection points of the families of curves $s(x,y)=0$ and $h(x,y)=k$ are found to lie on a hyperbola for a given $k$ in the range $k_{min} \leq k \leq k_{max}$, where the approximate values obtained are given by $k_{min}=-0.411$ and $k_{max}=0.347$. When a given $k<k_{min}$ and $k>k_{max}$, the intersection points are demonstrated to lie on an ellipse, and when $k=0$, intersection points lie on two straight lines passing through the origin. 
 
 We have also constructed the matrix polynomials denoted by $\Tilde{P}_{n}(z)$ for ($n=1,2,3$) by permuting the coefficient matrices of $P_{n}(z)$. The trace of $\Tilde{P}_{n}(z)$ is not real, which imposes an additional restriction to the points $z \in\mathbb{C}$, where $\Tilde{P}_{n}(z)$ can be PT-symmetric. This additional constraint thereby reduces the points $z \in\mathbb{C}$ where $\Tilde{P}_{n}(z)$ exhibits PT-symmetry. Along with other findings, we have also shown that the set of all solutions of the equations $s(x,y)=0$ and $h(x,y)=\pm k:k>0$, for a very large value of $k$ (of the order of $10^{17}$), associated with $P_{3}(z)$ have the property of being related to reflection across the line $y=x$. That is, by knowing the solutions $\{(x,y)\}$ of the equations $s(x,y)=0$ and $h(x,y)= k:k>0$ , the solutions of the equations $s(x,y)=0$ and $h(x,y)= -k:k>0$ can be obtained as $\{(y,x)\}$.
 The nature of PT-symmetry at the zeros of the matrix polynomials $P_{n}(z)$ and $\Tilde{P}_{n}(z)$ is also investigated. It has been observed that at the zeros, the matrix polynomial  $P_{n}(z)$ (for n=1,2,3) exhibits unbroken PT-symmetry, Whereas $\Tilde{P}_{n}(z)$ has no PT-symmetry at its zeros except for the case $n=1$.
 
 In the last part, we examine a trace-less matrix polynomial denoted by $Q_{10}(z)=z^{7}\,( P_3(z)-  \sigma_0)$ of degree $10$. By plotting the intersection points of the families of curves associated with $Q_{10}(z)$ defined by $s(x,y)=0$ and $h(x,y)=k$ for $k = -1,\, -0.5,\, 0,\, 0.5,\, 1$, we demonstrated that these intersection points lie on a figure closely resembling a slightly deformed ellipse for $k=-1$, $k=-0.5$, $k=0.5$ and $k=1$ whereas the intersection points lie on two straight lines passing through the origin for $k=0$. Similarly, for $k=\pm 0.0001$, some of the intersection points (16 points) of the families of curves are shown to lie on a slightly deformed ellipse, while other points (4 points) can be seen to lie on a different ellipse or two straight lines passing through the origin. This result may suggest that intersection points are likely to lie on an ellipse or a slightly deformed ellipse when the absolute value of $k$ is sufficiently large for matrix polynomial of the form $z^{m}\,( P_3(z)-  \sigma_0)$ for any $m \geq 1$. Meanwhile, for a small value of $k$, the intersection points may lie on different ellipses or deformed ellipses.

We will commence by outlining some fundamental definitions and notations.
The identity matrix  $\sigma_0=I$ and the Pauli matrices $\sigma_1,\sigma_2,\sigma_3$  together form a basis for $ M_2(\mathbb{C})$. Therefore, any matrix $H\in M_2(\mathbb{C})$  can be expressed as a linear combination of these matrices given by

\begin{equation}\label{Eq.(1.1)}
H = \sigma_{0}\,h^{0} + \pmb{\sigma \cdot h} = \left[\begin{array}{cc}
h^0 + h^3  & h^1 - ih^2 \\
h^1 + ih^2 & h^0 - h^3
\end{array}\right] 
\end{equation} 
where,
$h^0=h^0_R+ih^0_I, \,\pmb{\sigma}=(\sigma_1,\sigma_2,\sigma_3), \,
\mathrm{and} \,\,
\pmb{h}=\pmb{h}_R+i\pmb{h}_I =(h^1_R + ih^1_I, h^2_R + ih^2_I, h^3_R + ih^3_I)$, such that \,$\pmb{h}\in\mathbb{C}^3$ and \,$\pmb{h}_R,\pmb{h}_I\in\mathbb{R}^3$ . 
The characteristic polynomial of $H$ and its roots, respectively, are given by
\begin{eqnarray}
f(E) &=& E^2-2h^0E+(h^0)^2-\pmb{h}\cdot\pmb{h}\label{Eq.(1.2)}\hspace{2cm}
\\
 E&=&h^0\mp\sqrt{\pmb{h}\cdot\pmb{h}}\label{Eq.(1.3)}\hspace{2cm}
\end{eqnarray}
here $\pmb{h}\cdot\pmb{h}=\pmb{h}_R\cdot\pmb{h}_R-\pmb{h}_I\cdot\pmb{h}_I+2i \pmb{h}_R\cdot\pmb{h}_I$

If a Hamiltonian $H$  is PT-symmetric, it satisfies the condition $\left[H, PT\right] = 0$, where $ P $ represents the parity operator and $T$ is the time-reversal operator. For any $H\in M_2(\mathbb{C})$ , the commutation relation $\left[H, PT\right] = 0$ is equivalent to $H $ having a real characteristic polynomial \cite{Bender:2009mq}. Therefor from Eq. (\ref{Eq.(1.2)}), it can be concluded that  $H\in M_2(\mathbb{C})$  is PT-symmetric if and only if $ h^0_I = 0 $ and $\pmb{h}_R \cdot \pmb{h}_I = 0 $ \cite{Bender:2009mq}.

Definition~\cite{SINAP199627}:
Let $A_0,A_1,..,A_n \in \mathbb{C}^{p \times p}$ be n+1 square matrices with complex entries, and suppose that $A_n\neq0$, then 
\\
$P: \mathbb{C}$ --> $\mathbb{C}^{p \times p}$,
such that 
\begin{equation}\label{Eq.(1.4)}
P(z) = z^{n}\, A_n +  z^{n-1}\, A_{n-1} +..+ z\, 
 A_1 + A_0\hspace{1.1cm}
\end{equation}
is a matrix polynomial of degree $n$. The matrix polynomial $P(z)$ is a $p \times p$ matrix for which each entry is a polynomial in $z$ of degree at most $n$.
A complex number $z_0$ is said to be the zero of the matrix polynomial  $P(z)$ if   $P(z_0)$ is singular. If the leading coefficient $ A_{n}$ is non-singular, then the determinant of $P(z)$ denoted by $det[P(z)]$ is a polynomial of degree $np$. Thus a matrix polynomial $P$ of degree $n$  and $det(A_n)\neq0$ has $np$ zeros~\cite{SINAP199627}.

\section{ PT-Symmetry and Matrix Polynomials Formed by Pauli Matrices}
In this  section, we will analyse the matrix polynomials of the form 
\begin{equation}\label{Eq.(2.1)}
    P_{n}(z) = \Sigma^{n}_{j=0}\,\sigma_{j}\,z^{j} \hspace{4.3cm}
\end{equation}
for the cases $n=1,2,3$ where $n$ is the degree of the matrix polynomial, $\sigma_0$ is the identity matrix and $\sigma_{j}$,($j=1,2,3)$ are the Pauli matrices. We also study the matrix polynomials constructed by permuting the coefficient matrices of $P_{n}(z)$, which is denoted by $\tilde{P}_{n}(z)$. These matrix polynomials are not PT-symmetric in general. So we are interested in finding all $z\in\mathbb{C}$ where the matrix polynomial exhibits PT-symmetry. Since all $2\times 2$ PT-symmetric matrices are also pseudo hermitian \cite{Wang201322}, at the points where the matrix polynomials are PT-symmetric, it is also Pseudo Hermitian. For the following discussions, we introduce two functions, $s(x,y)$ and $h(x,y)$, to characterize the nature of PT-symmetry of a matrix polynomial. The functions are given by
\begin{align}
 s(x,y) &= \pmb{h}_R \cdot \pmb{h}_I \hspace{4cm}\label{Eq.(2.2)} \\
h(x,y) &= \pmb{h}_R \cdot \pmb{h}_R - \pmb{h}_I \cdot \pmb{h}_I \label{Eq.(2.3)} \hspace{4cm}
\end{align}

where $\pmb{h}_R$ and $\pmb{h}_I$ are the vectors associated with the matrix polynomial when expressed as 
Eq.(\ref{Eq.(1.1)}).  
By making use of Eq.(\ref{Eq.(1.3)}) the determinant of a $2\times 2 $ matrix polynomial can be expressed as 
\begin{equation}\label{Eq.(2.4)}
  det[P(z)] = (h^{0}_{R})^2 - (h^{0}_{I})^2 - h(x,y) + i\, 2( h^{0}_{R}  h^{0}_{I} - s(x,y)) \hspace{0.5cm}
\end{equation}
where $h^{0}_{R}$ and $h^{0}_{I}$ are functions of $x$ and $y$. For any $2\times2$ matrix polynomial, the determinant is a polynomial in $z$, and there exists at least one $z=(x,y)\in\mathbb{C}$ such that the determinant becomes real. Hence if the trace of a matrix polynomial is real, it exhibits PT-symmetry at least one point in the complex plane. More precisely, if the trace of a $2\times2$ matrix polynomial is real, then the matrix polynomial exhibits PT-symmetry at all  $z=(x,y)\in\mathbb{C}$ where $s(x,y)=0$, where its eigenvalues are given by $E= h^{0}_{R} \pm \sqrt{h(x,y)}$. In addition, at those points where $h(x,y)=k:k\geq0$, the matrix polynomial has real eigenvalues given by $E=h^{0}_{R} \pm \sqrt{k}$ (unbroken PT-symmetry). Similarly, at those points where $h(x,y)=k:k<0$, the matrix polynomial has complex pairs of eigenvalues given by $E=h^{0}_{R} \pm i \sqrt{|k|}$ (broken PT-symmetry). If the trace of a $2\times2$ matrix polynomial is not real, then the matrix polynomial exhibits PT-symmetry at all  $z=(x,y)\in\mathbb{C}$ where $s(x,y)=0$ and the imaginary part of the trace becomes zero, i.e. when $ h^0_I = 0 $. This additional requirement may reduce the number of points in the complex plane where the matrix polynomial possesses PT-symmetry. 
\subsection{PT-Symmetry in Matrix Polynomials of Degree 1}
1) Consider the  matrix polynomials of degree-1, given by 
 \begin{equation}\label{Eq.(2.5)}
P_{1}(z) = \Sigma^{1}_{j=0}\,\sigma_{j}\,z^{j}=\sigma_{0} + \pmb{\sigma} \cdot (z,0,0) = \left[\begin{array}{cc}
1 & z\\
z & 1
\end{array}\right]\hspace{0.6cm}
\end{equation}
The matrix polynomial $P_{1}(z)$ is generally not PT-symmetric. Since the trace of $P_{1}(z)$ is real, the PT-symmetric nature of $P_{1}(z)$  can be characterized by the functions given by
\begin{align}
    s(x,y) &= xy \hspace{5.4cm} \label{Eq.(2.6)}\\
    h(x,y) &= x^2 - y^2\hspace{5.4cm} \label{Eq.(2.7)}
\end{align}

 The determinant of $P_{1}(z)$ is given by 
 \begin{equation}
   \det[P_{1}(z)] = 1-z^2= 1-  h(x,y) - i\, 2 s(x,y) \label{Eq.(2.8)}\hspace{2.6cm}
 \end{equation}
 The matrix polynomial $P^{1}(z)$ is PT-symmetric at all points in the complex plane defined by the equation $s(x,y) = 0$. The points in the complex plane where the families of curves $ s(x,y) = 0 $ and $ h(x,y) = k:k>0 $ intersect are along the x-axis, whereas the points where the curves  $ s(x,y) = 0 $ and $ h(x,y) = k:k<0  $ intersect are along the y-axis. For a given $ k \neq0 $, the curves $ h(x,y) = k $ and $ s(x,y) = 0 $ intersect at two points, and the intersection points lie on hyperbola given by  Eq. (\ref{Eq.(2.7)}) as shown in Fig-1-(a) and Fig-1-(c). When $k=0$, the curves intersect at the origin, as shown in Fig-1-(b). Thus corresponding to the points on the x-axis $P_{1}(z)$ becomes a symmetric matrix and has real eigenvalues given by $E=1\pm x$ and corresponding to the points on the y-axis $P_{1}(z)$ becomes normal, with the property $[P_{1}(z)]^\dag P^{1}(z) = P_{1}(z) [P_{1}(z)]^\dag = \sigma_0 \det[P_{1}(z)]$,  and has complex conjugate pairs of eigenvalues given by $E=1\pm i y$. 
 The points $(\pm1,0)$, where the curves $s(x,y)=0$ and $h(x,y) =1$ intersect ( as shown in Fig-1-(c)), are the zeros of the matrix polynomials. The matrix polynomial exhibits unbroken PT-symmetry at the zeros.
 
 Note that the $\det[P_{1}(z^*)] = (\det[P_{1}(z)])^*$, and $\det[P_{1}(-z)]=\det[P_{1}(z)]$  i.e, the determinant has conjugate symmetry and even properties. Hence, the points where curves $s(x,y)=0$ and $h(x,y) =k$ intersect for a given $k$ will be symmetric about the x-axis or about the origin. 
 \begin{figure}[H]
    \centering
    \includegraphics[width=0.6\linewidth]{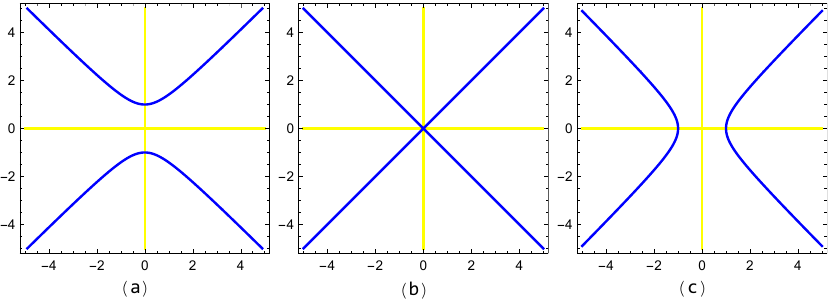}
    \caption{Intersections of the families of curves  $s(x,y) = xy = 0$ (Yellow) and $h(x,y)= x^2-y^2 = k$ (Blue) for the cases:  (a) $k = -1$, (b) $k = 0$, (c) $k = 1$}
    \label{fig:1}
\end{figure}

2) Now consider the matrix polynomial of degree 1, constructed by permuting the coefficient matrices of $P_{1}(z) $, given by
\begin{equation}\label{Eq.(2.9)}
\tilde{P}_{1}(z) = \sigma_{0}\, z + \pmb{\sigma} \cdot (1,0,0) = \left[\begin{array}{cc}
z & 1\\
1 & z
\end{array}\right]\hspace{2cm}
\end{equation}
The trace of $\tilde{P}_{1}(z)=2z$ is not real. In this case, the functions $s(x,y) $ and $h(x,y)$ are constant functions and always have the values $0$ and $1$ respectively. Hence, $\tilde{P}_{1}(z)$ becomes PT-symmetric when the trace of the matrix polynomial becomes real, i.e., along the x-axis where $\tilde{P}_{1}(z)$ is a symmetric matrix and has unbroken PT-symmetry with real eigenvalues $E=x\pm1$. The zeros of $\tilde{P}_{1}(z)$ coincide with the zeros of $P_{1}(z)$ i.e., the points $(\pm1,0)$, where $\tilde{P}_{1}(z)$ has unbroken PT-symmetry.

\subsection{PT-Symmetry in Matrix Polynomials of Degree 2}
1) Consider the matrix polynomial of degree 2 given by 
 \begin{equation}\label{Eq.(2.10)}
P_{2}(z)  = \Sigma^{2}_{j=0}\,\sigma_{j}\,z^{j}=\sigma_{0} + \pmb{\sigma} \cdot (z,z^2,0) = \left[\begin{array}{cc}
1 & z - i z^2\\
z + i z^2 & 1
\end{array}\right]\hspace{1cm}
\end{equation} 
The determinant of $P_{2}(z)$ is given by 
 \begin{equation}\label{Eq.(2.11)}
   \det[P_{2}(z)] = 1-z^2-z^4= 1-  h(x,y) - i\, 2 s(x,y)\hspace{1.5cm}
 \end{equation}
The matrix polynomial $P_{2}(z)$  is generally not PT-symmetric. The functions $s(x,y)$ and $h(x,y)$ associated with the matrix polynomial are given by 
\begin{align}
    s(x,y) &= xy + 2xy(x^2 - y^2) \label{Eq.(2.12)}\hspace{2cm} \\
    h(x,y) &= x^2 + (x^2 - y^2)^2 - (y^2 + 4x^2y^2) \hspace{2cm} \label{Eq.(2.13)}
\end{align}

The trace of the matrix polynomial $P_{2}(z)$  is real. Hence, the set of points in the complex plane for which $P_{2}(z)$ can be PT-symmetric is the same as the set of points where $s(x,y) = 0 $. 
For a given $k$, there are four points in the complex plane where the families of curves $ s(x,y) = 0 $ and $ h(x,y) = k $ of $P_{2}(z)$  intersect or the matrix polynomial exhibits PT-symmetry, except for the value of $ k = 0 $  and $k = -0.25$. For $ k = 0 $, there are three points where the families of curves $ s(x,y) = 0 $ and $ h(x,y) = k $ intersect or $P_{2}(z)$ exhibits PT-symmetry, and all three points lie on the y-axis, as shown in Fig-2-b. The intersections of the curves $ s(x,y) = 0 $ and $ h(x,y) = k $ for $ k = -1, 0, 1 $ are shown in Fig-2. The four zeros of the matrix polynomial are at the points defined by the intersection of the curves $s(x,y)=0$ and $ h(x,y) =1 $ as shown in Fig-2-(c). At the zeros of the matrix polynomial, $P_{2}(z)$ has unbroken PT-symmetry and has eigenvalues $E=0,2$.
\begin{figure}[H]
\centering
\includegraphics[width=0.7\linewidth]{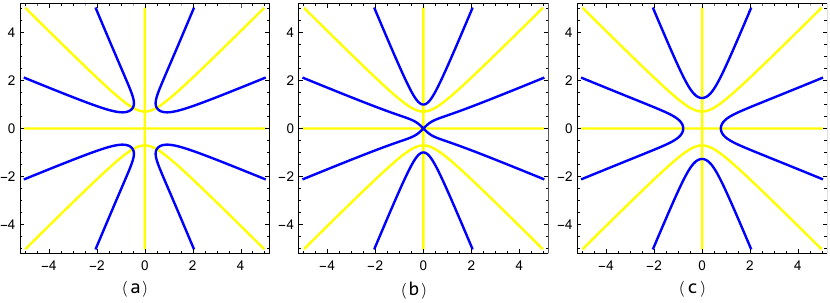}
\caption{Intersections of the families of curves $s(x,y) = xy + 2xy (x^2 -y^2) = 0$ (Yellow) and $h(x,y)= x^2 + (x^2 - y^2)^2 - (y^2 + 4 x^2 y^2) = k$ (Blue) for the cases: (a) $k = -1$, (b) $k = 0$, (c) $k = 1$}
\end{figure} \label{fig:2}

 At all points on the x-axis, $P_{2}(z)$  is Hermitian(unbroken PT-symmetry) with eigenvalues $E=1\pm x \sqrt{1+x^2}$; at all other points, $P_{2}(z)$  is non-hermitian and non-normal. Further, Eq.(\ref{Eq.(2.13)}) implies that for all points on the y-axis where $|y|<1$ and $|y|\neq0$, $h(x,y)<0$ and hence $P_{2}(z)$ must have broken PT-symmetry with eigenvalues $E=1\pm i y \sqrt{|y^2-1|}$ and for all points on the y-axis where $|y|>1$, $h(x,y)>0$ and therefore exhibit Unbroken PT-symmetry with eigenvalue $E=1\pm y \sqrt{y^2-1}$. For a given value of $ k< -0.25 $, the curves have no intersection on the x or y axis, as shown in Fig-3-(a) (for the range of $ -0.5 \leq k \leq -0.251$). When k = -0.25, we get the least number of points where the families of curves intersect, and these two points lie on the y-axis, as shown in Fig-3-(b). For a given $ -0.25 \leq k \leq0$, all 4  points lie on the y-axis, as shown in Fig-3-(c) (for the range of $ -0.249\leq k \leq 0$). 

\begin{figure}[H]
    \centering
    \includegraphics[width=0.8\linewidth]{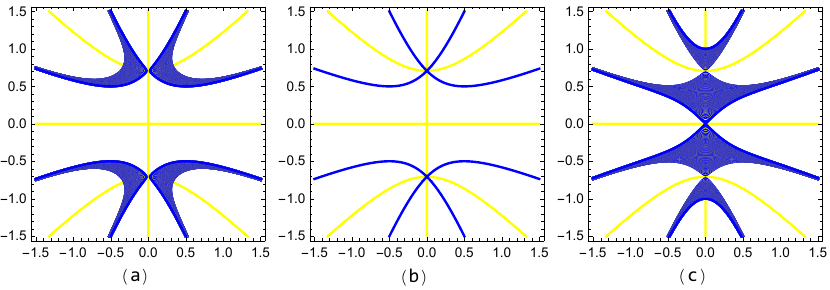}
    \caption{ Intersections of the families of curves $s(x,y) = xy + 2xy (x^2 -y^2) = 0$ (Yellow) and $h(x,y)= x^2 + (x^2 - y^2)^2 - (y^2 + 4 x^2 y^2) = k$ (Blue) for the cases: (a) $ -0.5\leq k\leq -0.251 $ (b) $k = -0.25$, (c) $ -0.249\leq k\leq0 $}
    \label{fig:3}
\end{figure}

Note that $\det[P_{2}(z)]$ has conjugate symmetry, and the property $\det[P_{2}(z)]=\det[P_{2}(-z)]$ (even). These properties of the determinant of $P_{2}(z)$ suggest that the solutions of the equations  $s(x,y)=0$ and $h(x,y)=k$ can lie on an ellipse or hyperbola or two straight lines passing through the origin or along a straight line. For a given value $ k< -0.25 $, the curves have no intersection on the x or y-axis. Therefore four intersection points can be seen lying on a circle because of conjugate symmetry and even properties of the determinant. That is, if $(x\neq0, y\neq0)$ is a solution, then the other three solutions must be \{(-x,-y),(-x,y),(x,-y)\}. Thus, when $ k< -0.25 $, all solutions must lie on a circle and at all points on the circle, $P_{2}(z)$ has broken PT-symmetry. 

For example by solving for the four points ${(x,y)}$ satisfying $s(x,y)=0$ and $h(x,y)=-1$ we get the points:

$\{(x,y),(-x,-y),(-x,y),(x,-y)\}$ where $x = \frac{1}{2}$ and $y=\frac{\sqrt{3}}{2}$.
All these points lie on the unit circle, as shown in Fig-4-a.
These points can also be seen as lying on an ellipse, a hyperbola, or two straight lines passing through the origin. If a solution lies on the y-axis, then conjugate property or even property of the determinant function guarantees one more solution symmetric to the x-axis. If a solution lies on the x-axis, then even property of the determinant function guarantees one more solution symmetric to the y-axis. Thus, if the curves intersect on the x-axis or y-axis, the remaining solutions need not lie on a circle. This is what happens for a given value of $k>0$; out of the four points where the families of curves $s(x,y)=0$ and $h(x,y)=k$ intersect, two points lie on the x-axis and the other two on the y-axis symmetrically, these points can be seen as lying on an ellipse. At all these points on the ellipse $P_{2}(z)$ has unbroken PT-symmetry.

For example, by solving for the four points 
$\{(x,y)\}$ 
satisfying 
$s(x,y)=0$ and $h(x,y)=1$ 
we get the points: 

\{\(
(0,y_1),(0,-y_1),(x_2,0),(-x_2,0)
\)\}, 
where $y_1=1.27202$, $x_2=0.786151$.  All these points lie on the ellipse with the length of the semi-major axis equal to $y_1$ and the length of the semi-minor axis equal to $x_2$, as shown in Fig-4-b.
\begin{figure}[H]
    \centering
    \includegraphics[width=0.6\linewidth]{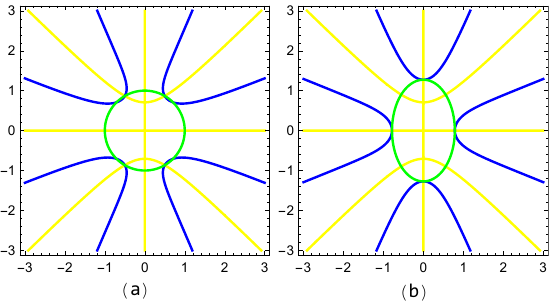}
 \caption{Intersections of the families of curves $s(x,y) = xy + 2xy (x^2 -y^2) = 0$ (Yellow) and $h(x,y)= x^2 + (x^2 - y^2)^2 - (y^2 + 4 x^2 y^2) = k$ (Blue) for the cases: (a) $k = -1$, (b) $ k=1$ }
\end{figure}\label{fig:4}

From the above discussion, we have the following realization about the matrix polynomial $P_{2}(z)$. Along all points on the x-axis, $P_{2}(z)$ is hermitian and has unbroken PT-symmetry. Along all points on the y-axis  where $|y|<1$ and $|y|\neq0$, $P_{2}(z)$  has broken PT-symmetry and for all points where $|y|>1$, $P_{2}(z)$  has Unbroken PT-symmetry. For a given $k>0$, the intersection points of the families of curves $s(x,y)=0$ and $h(x,y)=k$ lie on an ellipse. For a given $ -0.25 \leq k \leq0$, intersection points of the families of curves lie on the y-axis. Finally, for all $k<-0.25$, intersection points of the families of curves lie on a circle.

2) Now consider the matrix polynomial of degree 2, constructed by permuting the coefficient matrices of  $ P_{2}(z) $, given by 
\begin{equation}\label{Eq.(2.14)}
\tilde{P}_{2}(z)  = \sigma_{0} z + \pmb{\sigma} \cdot (z^2,1,0) = \left[\begin{array}{cc}
z & z^2 - i \\
z^2 + i  & z
\end{array}\right]\hspace{1cm}
\end{equation} 
The determinant of $\tilde{P}_{2}(z)$ is given by 
 \begin{equation}\label{Eq.(2.15)}
   \det[\tilde{P}_{2}(z)] = z^2-z^4 -1 = x^2 - y^2 - h(x,y) + i\, 2(xy-s(x,y))
 \end{equation}
The trace of $\tilde{P}_{2}(z)= 2 z $ is complex and the functions $s(x,y)$ and $h(x,y)$ associated with the matrix polynomial are given by 
\begin{align}
    s(x,y) &= 2xy (x^2 - y^2) \label{Eq.(2.16)} \hspace{3cm}\\
    h(x,y) &= (x^2 - y^2)^2 + 1 - 4x^2y^2 \label{Eq.(2.17)}\hspace{3cm}
\end{align}

The matrix polynomial $\tilde{P}_{2}(z)$ exhibits PT-symmetry when the trace becomes real i.e, $y=0$ and $s(x,y)=0$. Hence, $\tilde{P}_{2}(z)$  possesses PT-symmetry only along the x-axis, where it becomes Hermitian and has unbroken PT-symmetry with eigenvalues $E=x\pm \sqrt{x^4+1}$. The zeros of the matrix polynomials are defined by the intersection of the curves given by 
\begin{align}
    xy - s(x,y) &= 0 \label{Eq.(2.18)} \quad\text{and} \hspace{4cm}\\
    x^2 - y^2 - h(x,y) &= 0 \label{Eq.(2.19)}\hspace{4cm}
\end{align}

where $s(x,y)$ and $h(x,y)$ are given by Eq.(\ref{Eq.(2.16)}) and Eq.(\ref{Eq.(2.17)}) respectively.
At the four zeros of the matrix polynomial, $ \tilde{P}_{2}(z) $ is not PT-symmetric since the curves defined by Eq.(\ref{Eq.(2.18)}) and Eq.(\ref{Eq.(2.19)}) have no intersection on the x-axis as shown in the Fig-5.
\begin{figure}[H]
    \centering
    \includegraphics[width=0.35\linewidth]{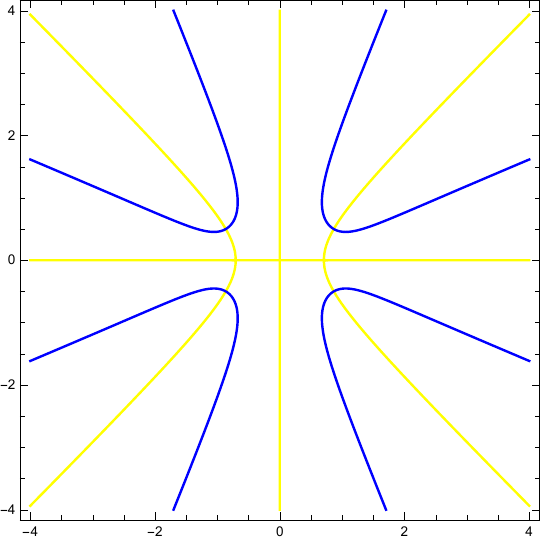}
    \caption{Zeros of $ \tilde{P}_{2}(z) $ defined by the intersection of the families of curves $xy - 2xy (x^2 -y^2) = 0$ (Yellow) and $x^2 - y^2 - ( (x^2 - y^2)^2 +1 - 4 x^2 y^2) = 0$ (Blue) }
\end{figure}\label{fig:5}

In this case, for a given $k\neq1$, the families of curves $ h(x,y)=k$ and $s(x,y)=0$ intersect at four points such that when $k<1$, there are no intersection points on the x-axis as shown in Fig-6-(a) and consequently no PT-symmetry and when $k>1$, two intersection points are on the x-axis symmetrically as shown in Fig-6-(c). When $k=1$, the curves intersect at one point (origin), as shown in Fig-6-(b). Thus, when the trace of a matrix polynomial is not real in general, the additional constraint on points in the complex plane where the trace of the matrix polynomial is real reduces the number of points where it can be PT-symmetric. 
\begin{figure}[H]
    \centering
    \includegraphics[width=0.75\linewidth]{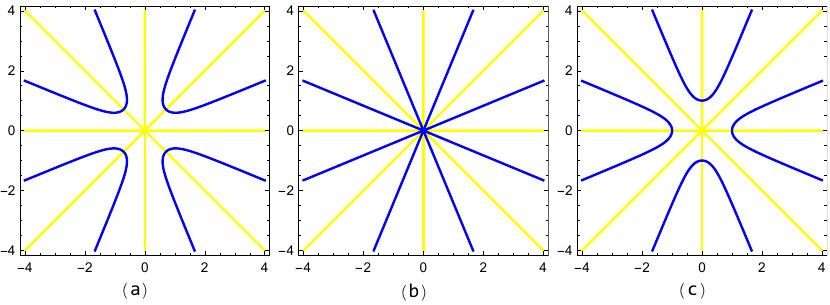}
    \caption{Intersections of the families of curves $s(x,y) = 2xy (x^2-y^2)  = 0$ (Yellow) and $h(x,y)= (x^2-y^2)^2 + 1 - 4 x^2y^2  = 0$ (Blue), for the cases: (a) $k = 0$, (b) $k = 1$,(c) $k = 2$}
\end{figure}\label{fig:6}

\subsection{PT-Symmetry in  Matrix Polynomials of Degree 3}

1) Consider the matrix polynomial of degree-3 given by 
\begin{equation}\label{Eq.(2.20)}
P_{3}(z) = \sigma_{0} + \pmb{\sigma} \cdot (z,z^2,z^3) = \left[\begin{array}{cc}
1 + z^3 & z - i z^2\\
z + i z^2 & 1 - z^3
\end{array}\right]
\end{equation} 
The determinant of $P_{3}(z)$ is given by 
 \begin{equation}\label{Eq.(2.21)}
   \det[P_{3}(z)] = 1-z^2-z^4-z^6= 1-  h(x,y) - i\, 2 s(x,y)\hspace{0.5cm}
 \end{equation}
The matrix polynomial $P_{3}(z)$ is not PT-symmetric for all $z$ values. The functions s(x,y) and h(x,y) associated with the matrix polynomials are given by
\begin{align}
    s(x,y) &= xy + 2xy(x^2 - y^2) + (x^3 - 3xy^2)(3x^2y - y^3) \label{Eq.(2.22)}\hspace{-2cm} \\
    h(x,y) &= x^6 - y^2 + y^4 - y^6 + x^4(1 - 15y^2) + x^2(1 - 6y^2 + 15y^4) \label{Eq.(2.23)}\hspace{-2cm}
\end{align}

The trace of matrix polynomial $P_{3}(z)$  is real, and hence it is PT-symmetric at all points where $s(x,y)=0$. The six zeros of the matrix polynomial are at the points defined by the intersection of $s(x,y) = 0$ and $h(x,y) = 1$, as shown in Fig-7-a; at these points, the matrix polynomial has unbroken PT-symmetry with eigenvalues $E=0,2$. At all points on the x-axis $P_{3}(z)$ is Hermitian, and at all other points, $P_{3}(z)$ is non-hermitian and non-normal. There are, at most, six points where the curves $s(x,y)=0)$ and  $h(x,y)=k:k\in\mathbb{R}$, for a given $k$ intersect. When $k =0$, two curves among the family of curves defined by the equation $h(x,y)=k$ merge at the origin, as shown in Fig-7-b, leading to the least number of points where $P_{3}(z)$  has PT-symmetry for a given k. Thus when $k=0$ there are 5 points where $s(x,y)=0$ and $h(x,y)=k$ intersect . The number of points where $P_{3}(z)$ is PT-symmetric is six except at $k=0$.  

\begin{figure}[H]
    \centering
    \includegraphics[width=0.8\linewidth]{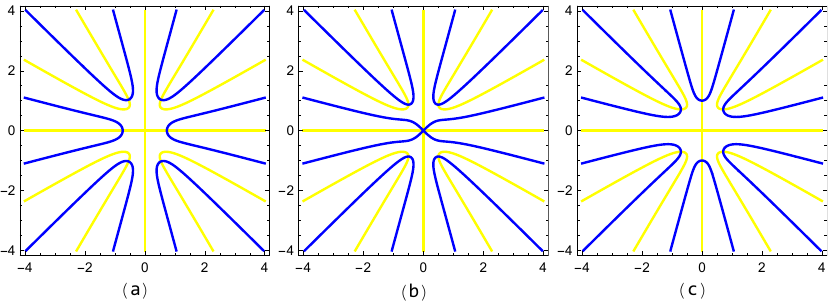}
    \caption{Intersections of the families of curves $s(x,y) = xy + 2xy(x^2-y^2) +(x^3 - 3 x y^2)(3x^2 y - y^3)=0$ (Yellow) and $ h(x,y) = x^6 - y^2 + y^4 -y^6 + x^4(1 - 15 y^2) + x^2(1-6y^2 + 15 y^4) =0$ (Blue), for the cases: (a) $k = 1$, (b) $k = 0$, (c) $k = -1$}
        \label{fig:7}
\end{figure}
As we discussed in the earlier case, $\det[P_{3}(z)]$ also has conjugate symmetry, and the property $\det[P_{3}(z)]=\det[P_{3}(-z)]$ (even). Hence, solutions of the equations  $s(x,y)=0$ and $h(x,y)=k$, for a given $k$, could lie on an ellipse or hyperbola or two straight lines passing through the origin or along a straight line. If there exists one solution $(x\neq0,y\neq0)$, then there must exist three other solutions given by $\{(-x,-y),(-x,y),(x,-y)\}$. Further, the remaining two solutions must lie on the x-axis or y-axis due to conjugate symmetry and even properties of $\det[P_{3}(z)]$.
The intersection points of the families of curves $s(x,y)=0$ and $h(x,y)=k$ for $k=0$ lie on two lines passing through the origin since one intersection point lies on the origin and the remaining four points are not on x-axis or y-axis. For a given value of $k$ in the range  $ 0<k\leq k_{max}$, where $k_{max}=0.347(approximately)$, four intersection points are given by $\{(x_1,y_1),(-x_1,-y_1),(-x_1,y_1),(x_1,-y_1): x_1\neq0, y_1\neq0\}$ and  the remaining two intersection points are given by $\{(x_2,0),(-x_2,0),: x\neq0, \}$, where $|x_2|<|x_1|$. Since $|x_2|<|x_1|$, these points lie on a hyperbola together. when the given value of $k>k_{max}$, four intersection points are given by $\{(x_1,y_1),(-x_1,-y_1),(-x_1,y_1),(x_1,-y_1): x_1\neq0, y_1\neq0\}$ and the remaining two intersection points are given by $\{(x_2,0),(-x_2,0),: x_2\neq0, \}$, where $|x_2|>|x_1|$. Since $|x_2|>|x_1|$, these points lie on an ellipse together. Similarly, for a given value of $k_{min}\leq k<0$, where $k_{min}=-0.411 ( approximately )$ the six intersection points lie on a hyperbola, and for a given $k<k_{min}$, the six intersection points lie on an ellipse. Thus, for a given $k_{min}\leq k \leq k_{max}:k\neq0$, intersection points lie on a hyperbola; otherwise, they lie on an ellipse. 

For example, the transition from intersection points lying on a hyperbola to those on an ellipse around $x_{min}$ and $x_{max}$ can be realized by solving the following equations for the numerical values of $x$ and $y$, where $s(x,y)$ and $h(x,y)$ are given by Eq.(\ref{Eq.(2.22)}) and Eq.(\ref{Eq.(2.23)}) respectively.

 \textbf Examples

 1) $s(x,y)=0$, $h(x,y)=-0.411$: Solving the  equations, we get the solutions given by

 $\{(x_1,y_1),(x_1,-y_1),(-x_1,y_1),(-x_1,-y_1),(0,y_2),(0,-y_2)\}$,

 where $x_1=0.565899$, $y_1=0.739613$ and $y_2=0.739209$.

 These points can lie on the hyperbola: $\frac{y^2}{b^2}-\frac{x^2}{a^2}=1$ with $b=y_2$. Since $b<y_1$, these points can not lie on an ellipse.

 The value of $a$ can be found using the equation $a^2= \frac{(x_1)^2 b^2}{(y_1)^2-b^2}$.

 2) $s(x,y)=0$, $h(x,y)=-0.412$: solving the  equations, we get the solutions given by

 $\{(x_1,y_1),(x_1,-y_1),(-x_1,y_1),(-x_1,-y_1),(0,y_2),(0,-y_2)\}$,

 where $x_1=0.566204$, $y_1=0.739425$ and $y_2=0.74005$.

 These points can lie on the ellipse: $\frac{x^2}{a^2} + \frac{y^2}{b^2} = 1$ with $b=y_2$. Since $ y_1 <b$, these points can not lie on a hyperbola.

 The value of $a$ can be found using the equation $a^2= \frac{(x_1)^2 b^2}{b^2-(y_1)^2}$. 

 3) $s(x,y)=0$, $h(x,y)=0.347$: solving the above equations, we get the solutions given by

 $\{(x_1,y_1),(x_1,-y_1),(-x_1,y_1),(-x_1,-y_1),(x_2,0),(x_2,0)\}$,

 where $x_1=0.51107$, $y_1=0.94431$ and $x_2=0.510938$.

 These points can lie on the hyperbola: $\frac{x^2}{a^2}-\frac{y^2}{b^2}=1$ with $a=x_2$. Since $a<x_1$, these points can not lie on an ellipse.

 The value of $b$ can be found by using the equation  $b^2= \frac{(y_1)^2 a^2}{(x_1)^2-a^2}$. 

 4) $s(x,y)=0$, $h(x,y)=0.348$: solving the above equations, we get the solutions given by

$\{(x_1,y_1),(x_1,-y_1),(-x_1,y_1),(-x_1,-y_1),(x_2,0),(x_2,0)\}$,

 where $x_1=0.511115$, $y_1=0.944488$ and $x_2=0.511504$.

 These points can lie on the ellipse: $\frac{x^2}{a^2} + \frac{y^2}{b^2} = 1$ with $a=x_2$. Since $ x_1 <a$, these points can not lie on a hyperbola.

 The value of $b$ can be found using the equation  $b^2= \frac{(y_1)^2 a^2}{a^2-(x_1)^2}$. 

The contour plots of $s(x,y)=0$ and $h(x,y) =k : -0.411 \leq k \leq0 $ and  $s(x,y)=0$ and $h(x,y) =k : 0\leq k \leq 0.347$ are shown 

in Fig-8-a and Fig-8-b respectively.
\begin{figure}[H]
    \centering
    \includegraphics[width=0.6\linewidth]{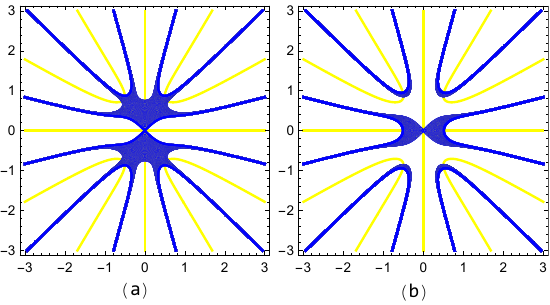}
    \caption{Intersections of the families of curves $s(x,y) = xy + 2xy(x^2-y^2) +(x^3 - 3 x y^2)(3x^2 y - y^3)=0$ (Yellow) and $ h(x,y) = x^6 - y^2 + y^4 -y^6 + x^4(1 - 15 y^2) + x^2(1-6y^2 + 15 y^4) =k$ (Blue), for the cases: (a) $ -0.411 \leq k \leq0 $, (b) $ 0\leq k \leq 0.347$}
        \label{fig:8}
\end{figure}
The Intersections of the families of curves, $s(x,y)=0$ and $h(x,y) =k$ for $k=1, 0.2, 0, -0.2, -1$, along with the corresponding intersection points lying on an ellipse with semi-major axis along the y-axis $\frac{x^2}{a^2} + \frac{y^2}{b^2}=1$ ( $ a=0.737353, 
 b=1.52$), a hyperbola intersecting the x-axis $\frac{x^2}{a^2} - \frac{y^2}{b^2}=1$  ($ a=0.409047,  b=1.05958$), two straight lines passing through the origin $y= \pm \sqrt{3}\,x$, a hyperbola intersecting the y-axis $-\frac{x^2}{a^2} + \frac{y^2}{b^2}=1$ ($ a=0.4042,  b=0.49544$), and the unit circle are shown in Fig-9.

\begin{figure}[H]
    \centering
    \includegraphics[width=0.9\linewidth]{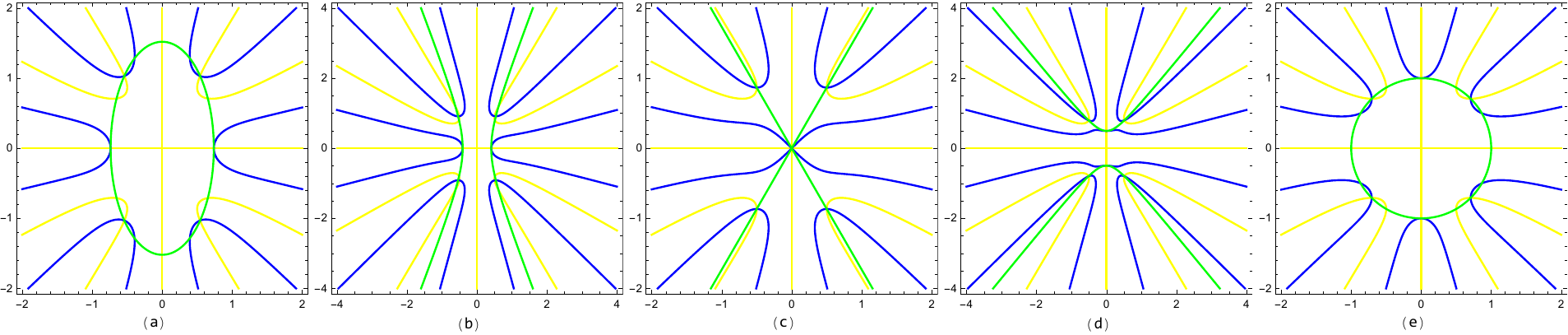}
    \caption{Intersections of the families of curves $s(x,y) = xy + 2xy(x^2-y^2) +(x^3 - 3 x y^2)(3x^2 y - y^3)=0$ (Yellow) and $ h(x,y) = x^6 - y^2 + y^4 -y^6 + x^4(1 - 15 y^2) + x^2(1-6y^2 + 15 y^4) =k$ (Blue), for the cases: (a) $  k =1 $, (b)$ k= 0.2$, (c) $ k= 0$, (d) $k=-0.2$, (e) $k=-1$}
        \label{fig:9}
\end{figure}

 From the above discussion, we have the following conclusions about the matrix polynomial $P_{3}(z)$.
 Along all points on the x-axis, $P_{3}(z), $ is Hermitian and has unbroken PT-symmetry, and along all points on the y-axis, it has broken Pt-symmetry. Further, for a given $k$,  the solution of the equations $s(x,y)=0$ and $ h(x,y)=k$ or, in other words, the intersection points of the families of curves $s(x,y)=0$ and $h(x,y)=k$ are found to lie on a hyperbola when  $k$ is in the range $k_{min} \leq k \leq k_{max}$, where the approximate values obtained are given by $k_{min}=-0.411$ and $k_{max}=0.347$. When a given $k<k_{min}$ and $k>k_{max}$, the intersection points lie on an ellipse, and when $k=0$, intersection points lie on two straight lines passing through the origin. Also, for a given $k$,  the solution of the equations $s(x,y)=0$ and $ h(x,y)=k$ or the intersection points of the families of curves defined by the equations lie on a hyperbola that intersects x-axis then at these intersection points  $P_{3}(z)$ exhibits unbroken PT-symmetry and if the intersection points lie on a hyperbola that intersects the y-axis, then at these points, $P_{3}(z)$ exhibits broken PT-symmetry.

2) Now consider the matrix polynomial of degree-3, constructed by permuting the coefficient matrices of $ P_{3}(z) $ given  by
\begin{equation}\label{Eq.(2.24)}
\tilde{P}_{3}(z) = \sigma_{0} z + \pmb{\sigma} \cdot (z^2,z^3,1) = \left[\begin{array}{cc}
z+1  & z^2 - i z^3\\
z^2 + i z^3 &  z -1
\end{array}\right]\hspace{1.1cm}
\end{equation} 
The determinant of $\tilde{P}_{3}(z)$ is given by 
 \begin{equation}\label{Eq.(2.25)}
   \det[\tilde{P}_{3}(z)] = z^2-z^4-z^6-1= x^2-y^2-  h(x,y) + 2i\,(xy-  s(x,y))\hspace{0.4cm}
 \end{equation}
 The trace of matrix polynomial $\tilde{P}_{3}(z)=2z$ is not real, which puts additional restriction on the points in the complex plane where the matrix polynomial can be PT-symmetric other than where $s(x,y)=0$. The functions $s(x,y)$ and $h(x,y)$ associated with the matrix polynomials are given by
\begin{align}
    s(x,y) &= 2xy(x^2 - y^2) + (x^3 - 3xy^2)(3x^2y - y^3)\label{Eq.(2.26)} \\
    h(x,y) &= 1 + x^6 + y^4 - y^6 + x^4(1 - 15y^2) + 3x^2y^2(-2 + 5y^2)\label{Eq.(2.27)}
\end{align}

The matrix polynomial $\tilde{P}_{3}(z)$ is PT-symmetric at the points where $y=0$ and $s(x,y)=0$, that is, along the x-axis. Thus, when $\tilde{P}_{3}(z)$ becomes PT-symmetric, it necessarily needs to be Hermitian.
The six zeros of the matrix polynomials are shown in Fig-10, where $x^2 - y^2
-h(x,y)=0$ and $xy-s(x,y)=0$.  In the case of $\tilde{P}_{3}(z)$, none of the zeros are on the x-axis; hence, at the zeros of the matrix polynomial, PT-symmetry cannot be observed.
\begin{figure}[H]
    \centering
    \includegraphics[width=0.3\linewidth]{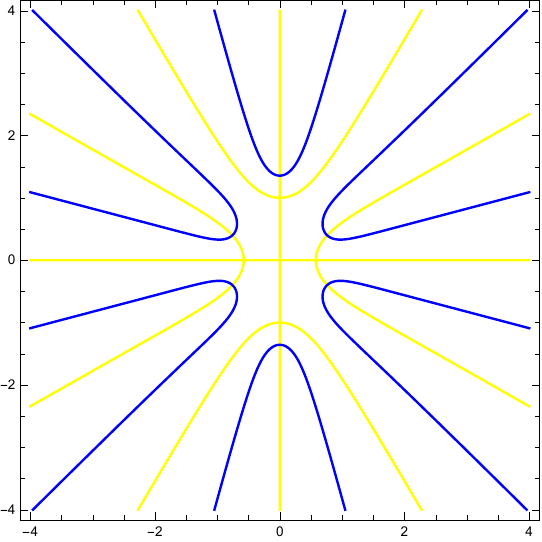}
    \caption{Zeros of $\tilde{P}_{3}(z)$ as intersections of the families of curves $xy-s(x,y)=0$ (Yellow) and $ x^2 - y^2
-h(x,y)=0$  (Blue)}
        \label{Fig:10}
\end{figure}
When the values of $k<1$, the function $h(x,y)=k$ have no intersection on the x axis as shown in Fig-11-(a) (for $0<k<1$) and hence $\tilde{P}_{3}(z)$ shows no PT-symmetry when $k<1$. The curve $h(x,y)=k$ intersects at one point on the x-axis (origin) when $k=1$ as shown in Fig-11-(b), and it intersects at two points on the x-axis symmetrically, when $k\geq1$ as shown in Fig-11-(c) (for k=2)  
 \begin{figure}[H]
    \centering
    \includegraphics[width=0.75\linewidth]{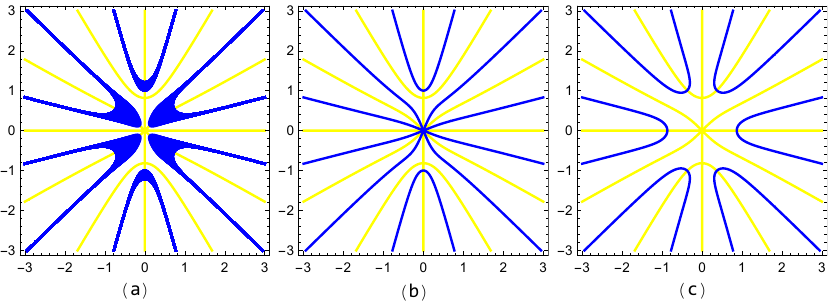}
    \caption{Intersections of the families of curves $s(x,y) = 2xy(x^2-y^2) +(x^3 - 3 x y^2)(3x^2 y - y^3)=0$ (Yellow) and  $h(x,y) = 1 + x^6 + y^4 - y^6 + x^4 (1 - 15 y^2) + 3 x^2 y^2 (-2 + 5 y^2)=k$ (Blue), for the cases: (a) $ 0 < k < 1$, (b)$ k= 1$, (c) $ k= 2$}
        \label{Fig:11}
\end{figure}

\subsubsection{Solutions of $s(x,y)=0$ and $h(x,y)=\pm k:k>0$, for very large value of k }

In the case of the matrix polynomials $P_{3}(z)$ , we examine  the solutions of the equations $s(x,y)=0$ and $h(x,y)=\pm k$, for very 

large value of $k=10^{17}$.

When $ h(x,y) = k = 10^{17}$, the solutions of the equations are given by:
$\{(x_1,y_1),(x_1,-y_1),(-x_1,y_1),(-x_1,-y_1),(x_2,0),(x_2,0)\}$

where $x_1=340.646$, $y_1=590.016$, and $x_2=681.292$. At all these points, the matrix polynomial has the eigenvalues given by 

$E=1\pm \sqrt{k}$.

When $ h(x,y) = k = - 10^{17}$, the solutions of the equations are given by:
$\{(y_1,x_1),(y_1,-x_1),(-y_1,x_1),(-y_1,-x_1),(0,x_2),(0,x_2)\}$.

At all these points, the matrix polynomial has the eigenvalues given by $E=1\pm i \sqrt{|k|}$.

The above results suggest that for very large values of k of order $10^{17}$, the  Solutions of $s(x,y)=0$ and $h(x,y)=\pm k:k>0$, have reflection property across the line $y=x$. That is, by knowing the solutions $\{(x,y)\}$ of the equations $s(x,y)=0$ and $h(x,y)= k:k>0$ , the solutions of the equations $s(x,y)=0$ and $h(x,y)= -k:k>0$ can be obtained as $\{(y,x)\}$.
 Thus if one set of solutions lies on the ellipses: $\frac{x^2}{a^2} + \frac{y^2}{b^2} = 1$, then the other set of solutions lies on the ellipse $\frac{x^2}{b^2} + \frac{y^2}{a^2} = 1$ for $h(x,y)=k$ and $h(x,y)=-k$ respectively. 
 Similar result has been observed for the case $\tilde{P}_{3}(z)$ also, for very large value of $k=10^{17}$. Since $\tilde{P}_{3}(z)$ exhibits PT-symmetry only for  $h(x,y)=k;k\geq1$, even if the reflection property exists  PT-symmetry can not be observed for $k=-10^{17}$.
\subsection{PT-Symmetry in matrix polynomial of degree 10}
Now we consider a matrix polynomial of degree-10, of the form  $z^{m}\,( P_3(z)-  \sigma_0)$ for $m=7$ given by 
\begin{equation}\label{Eq.(2.28)}
Q_{10}(z) =  \pmb{\sigma} \cdot (z^8,z^9,z^{10}) = \left[\begin{array}{cc}
 z^{10} & z^8 - i z^9\\
z^8 + i z^9 &  - z^{10}
\end{array}\right]
\end{equation} 
The determinant of $Q_{10}(z)$ is given by 
 \begin{equation}\label{Eq.(2.29)}
   \det[Q_{10}(z)] = -(z^{16}+z^{18}+z^{20}) = - ( h(x,y) + i\, 2 s(x,y))
 \end{equation}
The matrix polynomial $Q_{10}(z)$ is not PT-symmetric for all $z$ values. The functions s(x,y) and h(x,y) associated with the matrix polynomials are given by
\begin{equation}\label{Eq.(2.30)}
\begin{aligned}
    s(x,y) = &\quad (8 x^7 y - 56 x^5 y^3 + 56 x^3 y^5 - 8 x y^7)(x^8 - 28 x^6 y^2 + 70 x^4 y^4 - 28 x^2 y^6 + y^8) \\
             &\quad + (x^9 - 36 x^7 y^2 + 126 x^5 y^4 - 84 x^3 y^6 + 9 x y^8)(9 x^8 y - 84 x^6 y^3 + 126 x^4 y^5 - 36 x^2 y^7 + y^9) \\
             &\quad + (10 x^9 y - 120 x^7 y^3 + 252 x^5 y^5 - 120 x^3 y^7 + 10 x y^9)(x^{10} - 45 x^8 y^2 + 210 x^6 y^4 - 210 x^4 y^6 + 45 x^2 y^8 - y^{10}) \\
    h(x,y) = &\quad x^{20} + x^{18} (1 - 190 y^2) + x^2 y^{14} (-120 + 153 y^2 - 190 y^4) + y^{16} (1 - y^2 + y^4) \\
             &\quad - 286 x^{10} y^6 (28 - 153 y^2 + 646 y^4) - 60 x^{14} y^2 (2 - 51 y^2 + 646 y^4) \\
             &\quad + 5 x^4 y^{12} (364 - 612 y^2 + 969 y^4) + 78 x^8 y^8 (165 - 561 y^2 + 1615 y^4) \\
             &\quad + 26 x^{12} y^4 (70 - 714 y^2 + 4845 y^4) + x^{16} (1 - 153 y^2 + 4845 y^4) \\
             &\quad - 4 x^6 y^{10} (2002 - 4641 y^2 + 9690 y^4).
\end{aligned}
\end{equation}
Since $Q_{10}(z)$  is traceless, it is PT-symmetric at all points where $s(x,y)=0$. The family of curves $s(x,y)=0$ and $h(x,y)= k:k\neq0$ intersect at 20 points and when $k=0$ intersection points are reduced to 5. The eigenvalues of the matrix polynomial at the intersection points are given by $E=\pm \sqrt{k}$ and $E=\pm i \sqrt{|k|}$, for $k>0$ and $K<0$, respectively. When $|k|\geq 1$, the intersection points lie on a slightly deformed ellipse as shown in Fig-12-(a) (for $k =-1$) and Fig-12-(c) (for $k=1$). When $k=0$, the curves intersect  at the points  given by $\{(x,y),(x,-y),(-x,y),(-x,-y),(0,0)\}$: where $x=\frac{1}{2}$ and $y=\frac{\sqrt{3}}{2}$ as shown in Fig-12-(b). Around the value of $|k|=0.5$, all the intersection points are observed to lie on a slightly deformed ellipse, as shown in Fig-13-(a)(for k= -0.5) and Fig-13-(b)(for k=0.5). As the value of $|k|$ approaches zero, the intersection points are observed to not lie on a single ellipse or a single deformed ellipse; instead, 4 points lie on one ellipse, and the remaining 16 points are shown to lie on another deformed ellipse, as shown in Fig-13-(c)(for k=-0.0001) and Fig-13-(d)(for k=0.0001). This structure continues until $k=0$, and when $k=0$, the five intersection points lie on two straight lines passing through the origin. 
  \begin{figure}[H]
    \centering
    \includegraphics[width=0.7\linewidth]{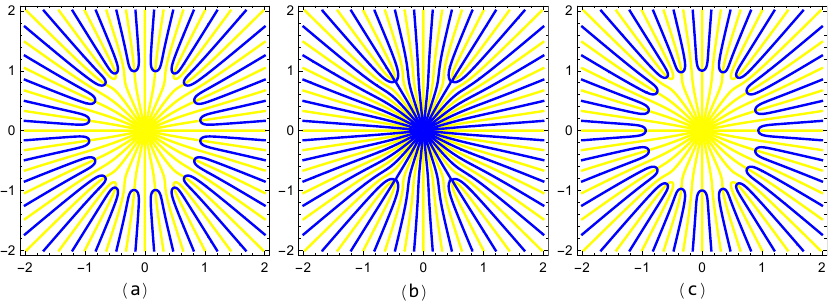}
    \caption{Intersections of the families of curves $s(x,y)$(Yellow) and $h(x,y)=k$(Blue), of $Q_{10}(z)$ for the cases: (a) $ k = -1$, (b) $ k= 0$ (c) $ k= 1$}
        \label{Fig:12}
\end{figure}
 \begin{figure}[H]
    \centering
    \includegraphics[width=0.9\linewidth]{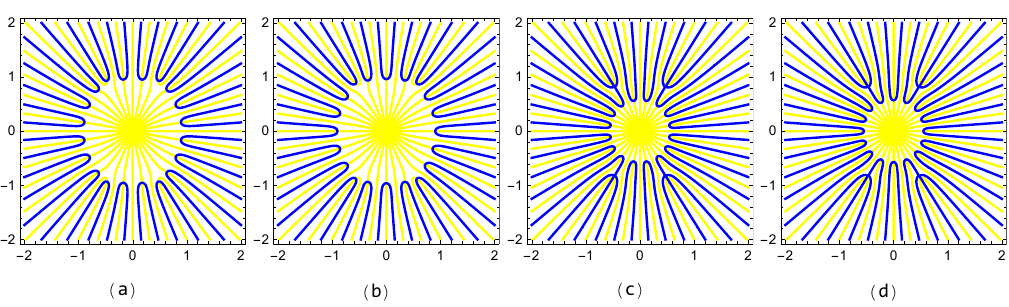}
    \caption{Intersections of the families of curves  $s(x,y)$(Yellow) and $ h(x,y)=k$ (Blue), of $Q_{10}(z)$  for the cases: (a) $  k = -0.5 $, (b) $ k= 0.5$, (c) $ k= -0.0001$, (d) $ k= 0.0001$}
        \label{Fig:13}
\end{figure}

This result may suggest that for matrix polynomial of the form $z^{m}\,( P_3(z)-  \sigma_0)$, all the intersection points are likely to lie on an ellipse or a slightly deformed ellipse when the absolute value of k is sufficiently large (i.e., not close to zero) for any $m \geq1$. In contrast, for small values of k, the intersection points are likely to lie on different ellipses or deformed ellipses. 
\section{Summary}\label{sec:3}

$2\times2$ matrix polynomials of the form  $P_{n}(z)= \Sigma^{n}_{j=0}\,\sigma_{j}\,z^{j}$ for $n= 1, 2, 3$ are examined for their PT-symmetric nature in the complex plane. The points where such matrix polynomials exhibit broken PT-symmetry and unbroken PT-symmetry are analyzed. It is demonstrated that these points lie on geometric figures such as hyperbolas, ellipses, two lines passing through the origin or straight lines. 

Furthermore, the matrix polynomial $P_3(z)$, for a given $k:k>0$ and sufficiently large, the points where eigenvalues of the matrix polynomial equal to $1 \pm \sqrt{k}$ and the points where eigenvalues equal to $1 \pm i \sqrt{k} $ are shown to related by reflection across the line $y=x$. The matrix polynomial $P_n(z)$ for the cases $n=1,2,3$ is found to have unbroken PT-symmetry at all of their zeros. 

Matrix polynomial $\tilde{P}_n(z)$ are constructed by permuting the coefficient matrices of $P_n(z)$ for $n=1,2,3$ and are found to not possess PT-symmetry at the zeros of the matrix polynomials except for $n=1$. A matrix polynomial of degree 10 of the form $z^{m}\,( P_3(z)-  \sigma_0)$ is also examined, leading to the speculation that the points where it can have eigenvalues $\pm \sqrt{k}$ or $\pm i \sqrt{k}$ (with k  positive and sufficiently large) lie on an ellipse or a slightly deformed ellipse for any $m \geq 1$. When the value of k is close to zero, the points may lie on different ellipses or deformed ellipses.

\section*{Acknowledgments}\label{sec:4}
The first author (Stalin Abraham) acknowledges financial support from AFL Pvt. Ltd. through the Cyrus Guzder Fellowship. I would like to express my sincere gratitude to Arnab Goswami and Dr Sudhir R. Jain for their helpful discussions.

\section*{References}
\bibliographystyle{unsrt}
\bibliography{aipsamp}
\end{document}